\documentclass[twocolumn]{aastex62}

\usepackage{wasysym}
\usepackage{savesym}
\savesymbol{tablenum}
\usepackage{enumitem}
\setlist[description]{leftmargin=\parindent,labelindent=\parindent}
\usepackage{siunitx}
\usepackage{multirow}
\restoresymbol{SIX}{tablenum}
\received{12 July, 2023}
\revised{n/a}
\accepted{n/a}
\submitjournal{ApJ}

\shorttitle{Survey of Moon-Forming Impacts I}
\shortauthors{Timpe et al.}
\begin{document}

\title{A Systematic Survey of Moon-Forming Giant Impacts I: Non-rotating bodies}

\correspondingauthor{Miles Timpe}
\email{mtimpe@proton.me}

\author[0000-0003-1938-7877]{Miles Timpe}
\affil{Institute for Computational Science \\
University of Z{\"u}rich \\
Winterthurerstrasse 190 \\
8059 Z{\"u}rich, Switzerland}

\author[0000-0002-4535-3956]{Christian Reinhardt}
% \email{christian.reinhardt@ics.uzh.ch}
\affil{Institute for Computational Science \\
University of Z{\"u}rich \\
Winterthurerstrasse 190 \\
8059 Z{\"u}rich, Switzerland}

\author[0000-0001-9682-8563]{Thomas Meier}
% \email{thomas.meier5@uzh.ch}
\affil{Institute for Computational Science \\
University of Z{\"u}rich \\
Winterthurerstrasse 190 \\
8059 Z{\"u}rich, Switzerland}

\author[0000-0001-7565-8622]{Joachim Stadel}
\affil{Institute for Computational Science \\
University of Z{\"u}rich \\
Winterthurerstrasse 190 \\
8059 Z{\"u}rich, Switzerland}

\author[0000-0001-5996-171X]{Ben Moore}
\affil{Institute for Computational Science \\
University of Z{\"u}rich \\
Winterthurerstrasse 190 \\
8059 Z{\"u}rich, Switzerland}



\begin{abstract}
In the leading theory of lunar formation, known as the giant impact hypothesis, a collision between two planet-size objects resulted in a young Earth surrounded by a circumplanetary debris disk from which the Moon later accreted. The range of giant impacts that could conceivably explain the Earth-Moon system is limited by the set of known physical and geochemical constraints. However, while several distinct Moon-forming impact scenarios have been proposed---from small, high-velocity impactors to low-velocity mergers between equal-mass objects---none of these scenarios have been successful at explaining the full set of known constraints, especially without invoking controversial post-impact processes. In order to bridge the gap between previous studies and provide a consistent survey of the Moon-forming impact parameter space, we present a systematic study of simulations of potential Moon-forming impacts. In the first paper of this series, we focus on pairwise impacts between non-rotating bodies. Notably, we show that such collisions require a minimum initial angular momentum budget of approximately $2~J_{EM}$ in order to generate a sufficiently massive protolunar disk. We also show that low-velocity impacts ($v_{\infty} \lesssim 0.5~v_{esc}$) with high impactor-to-target mass ratios ($\gamma \to 1$) are preferred to explain the Earth-Moon isotopic similarities. In a follow-up paper, we consider impacts between rotating bodies at various mutual orientations.
\end{abstract}


\section{Introduction} \label{sec:intro}

The prevailing theory of lunar formation is known as the Giant Impact Hypothesis, which posits that Earth's Moon is the result of an early and energetic impact event between two planetary-size bodies \citep{hartmannSatelliteSizedPlanetesimalsLunar1975, cameronOriginMoon1976}. In the leading version of this theory, which is generally referred to as the ``canonical'' Moon-forming impact, the young Earth suffered an oblique and relatively low-velocity impact by a Mars-sized impactor. This class of impacts corresponds to an impactor-to-target mass ratio of $\gamma \simeq 0.1$ and an impact velocity of $v_{imp} \simeq v_{esc}$, where $v_{esc}$ is the mutual escape velocity of the colliding bodies. Early simulations suggested that the canonical scenario could place approximately one lunar mass of material into orbit in the form of a circumplanetary disk while simultaneously reproducing the angular momentum budget of the Earth-Moon system and the low iron content of the Moon \citep{canupOriginMoonGiant2001, canupSimulationsLateLunarforming2004}. 

However, since the canonical scenario was proposed, improved constraints on the Earth-Moon system and advances in simulation techniques have brought the canonical scenario under renewed scrutiny. A well understood short-coming of the canonical scenario is its inability to explain the remarkable isotopic similarity of Earth's mantle and lunar samples returned by the Apollo missions. Indeed, the isotopic composition of the lunar and terrestrial mantles are indistinguishable when measured for several isotope ratios, including $^{18}O/^{17}O$ \citep{wiechertOxygenIsotopesMoonForming2001}, $^{50}Ti/^{47}Ti$ \citep{zhangProtoEarthSignificantSource2012}, and $^{182}W/^{184}W$ \citep{touboulLateFormationProlonged2007}. 

In the context of giant impact simulations, these measurements have significantly constrained the post-impact compositional difference between the Earth and the protolunar disk. This is a problem for the canonical scenario because most of the material that ends up in the impact-generated disk is derived from the impactor \citep{canupSimulationsLateLunarforming2004}. The extent to which this translates to differences in the isotopic fingerprints of the Earth and protolunar disk depends on the pre-impact isotopic compositions of the colliding bodies. If the difference between the pre-impact isotopic fingerprints of the target and impactor is large, then even a small difference in the post-impact compositions of the Earth and the protolunar disk will result in significant isotopic differences. In contrast, if the impactor has the same pre-impact isotopic composition as the target, then a preponderance of impactor material in the disk is not a problem. However, this would imply that the target and impactor formed at a similar heliocentric distance in the protoplanetary disk.  While theoretically possible, simulations of the terrestrial planet formation that track the origin of the accreted bodies show that such a scenario is a low-probability event (\citealt{kaibFeedingZonesTerrestrial2015,kaibBriefFollowupRecent2015}). 

Another way in which the canonical scenario might be reconciled with distinct isotopic fingerprints of the colliding bodies is through a post-impact mixing process that equilibrates Earth's magma ocean with the inner edge of the protolunar disk \citep{pahlevanEquilibrationAftermathLunarforming2007}. However, such processes require long time scales and therefore imply a formation time for the Moon that is orders of magnitude longer than that predicted by N-body simulations of lunar accretion  \citep{idaLunarAccretionImpactgenerated1997, kokuboEvolutionCircumterrestrialDisk2000}. This complication is even more pronounced for heavier (i.e., more refractory) elements such as titanium, which has also been shown to be indistinguishable between the Earth's mantle and the Moon \citep{zhangProtoEarthSignificantSource2012}.

Nevertheless, recent numerical studies have strengthened the case for post-impact equilibration. For example, post-impact mixing between material derived from the target and impactor may have been substantially underestimated because the different flavors of the Smoothed Particle Hydrodynamics (SPH) method (e.g., \citealt{springelCosmologicalSmoothedParticle2002}) used in most giant impact simulations suppress mixing \citep{agertzFundamentalDifferencesSPH2007, dengEnhancedMixingGiant2019}. Furthermore, \citet{dengEnhancedMixingGiant2019} showed that using a numerical method more suitable for investigating mixing substantially increases the amount of target material placed into orbit under the canonical scenario. \citet{dengPrimordialEarthMantle2019} additionally demonstrated that the canonical scenario is both consistent with the known isotopic constraints and successfully reproduced the heterogeneity of Earth's mantle, showing that it is a natural consequence of such a collision.

Whereas the canonical scenario generally assumes that the Moon accretes out of the protolunar disk, a recent study by \citet{kegerreisImmediateOriginMoon2022} shows that the Moon could instead be formed by the gravitational collapse of the outermost region of the arm-like structure that is observed in the canonical scenario. During repeated tidal encounters with the post-impact Earth, the Moon can accrete a thick layer of mantle material which could explain the isotopic similarity if such a vertical stratification can remain until today. However, it is unclear to which extent the gravitational collapse of impactor material is enhanced due to numerical issues and if such a vertical stratification can persist over longer time scales or if it will be erased by long-term geological processes.

Finally, recently discovered differences in vanadium isotopes between Earth and lunar samples \citep{nielsenIsotopicEvidenceFormation2021} suggest that these differences can only be explained by differences in the pre-impact bodies' differentiation processes. The vanadium isotope measurements are therefore inconsistent with equilibration after an impact, implying that any impactor would have had to have been isotopically very similar to Earth. In such a scenario, the Moon is formed mostly from impactor material as predicted by the canonical scenario.

Alternatively, several novel impact scenarios have been proposed which are notably distinct from the canonical scenario. The most successful of these are high impact energy and high angular momentum scenarios in which near-perfect mixing is achieved, either due to merging of similar mass embryos \citep{canupFormingMoonEarthlike2012} or ejection of mantle material from a rapidly spinning proto-Earth hit by a very small impactor at about three times the mutual escape velocity \citep{cukMakingMoonFastSpinning2012}. Furthermore, such impacts can result in the formation of circumplanetary structures known as ``synestias'' which allow the young protolunar disk to continue exchanging material with the Earth's mantle following the impact \citep{lockOriginMoonTerrestrial2018}, further enhancing post-impact equilibration.

However, simulations that investigate the giant-impact stage during terrestrial planet formation show that equal mass collisions are very rare (\citealt{jacobsonLunarTerrestrialPlanet2014, kaibFeedingZonesTerrestrial2015, kaibBriefFollowupRecent2015}) and would occur very early. Such equal mass collisions could therefore be difficult to reconcile with the age of the Moon \citep{jacobsonHighlySiderophileElements2014}. Likewise, the large impact velocities required by \citet{cukMakingMoonFastSpinning2012} are not observed in the simulations of \citet{kaibFeedingZonesTerrestrial2015} and the large pre-impact rotation of the proto-Earth suggests that it experienced a similar-mass merger before the Moon-forming impact, which is a very rare event according to the same study. 

The largest challenge for such high angular momentum models is to explain how the excess angular momentum of approximately 1-2 $J_{EM}$ can be lost in order to be consistent with observations. \citet{cukMakingMoonFastSpinning2012} proposed that evection resonances with the Sun could remove the required amount of angular momentum but recent studies suggest that the parameter space for which this mechanism is effective is narrow and its efficiency is strongly dependent on the choice of tidal model \citep{wardAnalyticalModelTidal2020, rufuTidalEvolutionEvection2020}. Thus, a particular difficulty of lunar formation theory continues to be the identification of a giant impact scenario that can simultaneously reproduce the angular momentum budget of the Earth-Moon system and the isotopic similarity of the Moon and Earth's mantle. 

In addition to concerns about the angular momentum and isotopic constraints, the canonical Moon-forming impact also struggles to produce sufficiently massive protolunar disks. Indeed, the simulations presented here show that the class of canonical Moon-forming impacts cannot produce disks with sufficient masses to explain the current mass of the Moon, much less than the more massive disks that appear to be required by accretion studies. This is a significant problem because N-body simulations of accretion in the protolunar disk suggest that accretion rates are in the range of 10-55\% and therefore a disk mass of at least two lunar masses is required to form the Moon \citep{kokuboEvolutionCircumterrestrialDisk2000}. 

Thus, as it stands, several distinct Moon-forming impact scenarios have been proposed that---sometimes necessarily in combination with post-impact processes---are capable of reproducing certain constraints of the Earth-Moon system. However, to date, none of these individual scenarios, either with or without effective post-impact processes, are capable of reproducing all necessary constraints of the Earth-Moon system. We further note that prior work investigating Moon-forming impacts has largely been focused on explaining specific observational constraints and were therefore limited in their range of pre-impact parameters that they considered. Notably, with the exception of \citet{canupLunarformingCollisionsPreimpact2008} and \citet{ruiz-bonillaEffectPreimpactSpin2021}, pre-impact rotation of the target and impact was neglected in such studies.

The purpose of the present study is therefore to provide the community with a comprehensive survey of the parameter space of potential Moon-forming impacts and provide a systematic analysis of the collision outcomes. The simulations in this study assume a single giant impact event and the subsequent post-impact analysis determines whether any such event can simultaneously explain the observed physical, compositional, and geochemical constraints of the Earth-Moon system. We have chosen to split the results into two papers in order to keep the results tractable. The present paper (hereafter Paper I) focuses on the subset of collisions \emph{without} pre-impact rotation. The follow-up paper (hereafter Paper II) considers collisions \emph{with} pre-impact rotation of the target and impactor for a wide range of rotational configurations (e.g., co-rotating, counter-rotating). Paper I is intended to serve as a baseline for understanding the results of the rotating impacts discussed in Paper II.

\section{Constraints on post-impact properties} \label{sec:constraints}

There are a number of empirically determined constraints on the Earth-Moon system that must be met in order for a Moon-forming simulation to be considered successful (see \citealt{canupDynamicsLunarFormation2004, barrOriginEarthMoon2016} for a modern and comprehensive review). These constraints are: the total angular momentum budget of the Earth-Moon system ($J_{EM}$), the protolunar disk mass ($M_{d}$) as a proxy for the mass of the Moon ($M_{\leftmoon}$), the iron fraction of the post-impact Earth ($F^{Fe}_{\oplus}$), the iron fraction of the protolunar disk ($F^{Fe}_{d}$) as a proxy for the iron fraction of today's Moon ($F^{Fe}_{\leftmoon}$), and the difference in impactor-to-target material between the planet and disk ($\delta_{pd}$) as a proxy for the isotopic similarity of the Earth and Moon. There are other physical properties of the Earth-Moon system which do not strictly need to be explained by the simulations. These properties can readily be explained by, for example, post-impact dynamical processes, such as the inclination of the lunar orbit ($\theta_{\leftmoon}$), and are therefore not considered in this work.

\subsection{Post-impact angular momentum budget} \label{sec:constraint-am}

The constraint on the post-impact angular momentum budget is set by the current angular momentum of the Earth-Moon system and any subsequent processes which could conceivably alter the angular momentum of the system following the impact. Currently, the only known process by which a significant amount of angular momentum could have been removed from the Earth-Moon system is the Solar Evection Resonance (SER) \citep{cukMakingMoonFastSpinning2012}, which transfers angular momentum from the Earth-Moon system to the Earth's heliocentric orbit. 

The amount of angular momentum that can be drained from the Earth-Moon system in this way is still debated and the results depend strongly on the underlying tidal model \citep{rufuTidalEvolutionEvection2020}. Some studies have suggested that no more than a few percent of the initial post-impact angular momentum can be lost through the SER, whereas other studies have suggested that up to 2-3 $J_{EM}$ can be removed in this way. While the SER could theoretically remove a significant fraction of the post-impact angular momentum, it makes sense to favor impact scenarios with post-impact angular momenta as close to $J_{EM}$ as possible. This reduces the reliance on the SER, which is not well understood and requires post-impact dynamical configurations that are difficult to achieve. Thus, while we consider impact scenarios that produce post-impact angular momentum budgets of up to several $J_{EM}$, given two successful scenarios in which only the post-impact angular momenta differed, it would be reasonable to favor the scenario with a total post-impact angular momentum budget closer to $J_{EM}$.

\subsection{Protolunar disk mass} \label{sec:constraint-disk-mass}

The mass and composition of the protolunar disk is used as a proxy for the mass and composition of the Moon, which is assumed to form at a later time via accretion from the disk \citep{salmonLUNARACCRETIONROCHEINTERIOR2012,idaLunarAccretionImpactgenerated1997}. The SPH simulations used to study giant impacts cannot subsequently follow the lunar accretion process, as the timescale is orders of magnitude longer and the computational cost therefore prohibitive. Nevertheless, studies of the subsequent accretion process (decoupled from the impact simulations) have been carried out using numerical techniques designed specifically for that purpose \citep{takedaAngularMomentumTransfer2001,lockOriginMoonTerrestrial2018,nakajimaInvestigationInitialState2014}. These studies indicate that less than half of the protolunar disk material ends up in the Moon \citep{kokuboEvolutionCircumterrestrialDisk2000, salmonLUNARACCRETIONROCHEINTERIOR2012}. This suggests that the post-impact disk must contain at least $2~M_{\leftmoon}$ worth of material to subsequently form a body with a mass at least that of the Moon. Similarly, too massive a disk would presumably result in several moons or a Moon that is much too massive. However, no upper limit has been systematically determined. Nevertheless, it is reasonable to assume that the protolunar disk cannot be much more massive than a few lunar masses given typical accretion efficiencies.

\subsection{Iron content of the Earth and protolunar disk} \label{sec:constraint-iron}

Following an impact, iron from the cores of the target and impactor will be distributed between the post-impact Earth, the protolunar disk, and the ejecta. Iron located in the protolunar disk may be incorporated into the Moon as it accretes from the protolunar disk material. The amount of iron that ends up in the protolunar disk and is subsequently accreted into the Moon (assuming some accretion efficiency) must match the constraints derived from measurements and assumed lunar geological processes. Given our current understanding of the iron content of Earth, this means that roughly $0.33~M_{\oplus}$ of iron should end up in the post-impact Earth. 

In the case of the Moon, studies have constrained its iron fraction to be $\leq$~1.5\% \citep{williamsLunarInteriorProperties2014}, meaning that any successful impact scenario will have to avoid injecting any significant amount of iron into the protolunar disk. In this respect, it is difficult to set a hard upper limit on the iron fraction of the protolunar disk due to the unknown accretion efficiency of iron into the Moon. Thus, while the Moon is constrained to $\leq$~1.5~\% iron by mass, the constraint for the protolunar disk is likely higher because some of the iron may not be accreted into the Moon and may instead be ejected or reaccreted by the Earth. The initial radial distribution of the iron in the protolunar disk will likely play an important role as well, given that the iron inside and outside of the Roche limit will be subject to different dynamical processes. The long-term evolution the protolunar disk, however, is beyond the scope of this work. Future studies are needed to constrain the accretion efficiency of iron and set an upper limit and distributional constraints on the iron in the protolunar disk. 

\subsection{Compositional similarity of Earth and the protolunar disk} \label{sec:constraint-mixing}

Since the first lunar samples were returned by the Apollo missions, it has been clear that the Moon and the Earth---or at the very least their mantles---exhibit a remarkable geochemical similarity. More modern studies have uncovered additional isotopic similarities across several elements. However, simulations of giant impacts are not capable of tracking isotopic ratios directly and, as a result, the relative fraction of impactor material between Earth and the protolunar disk is used as a proxy,

\begin{equation} \label{eq:delta-mix}
    \delta_{pd} = \left(\frac{N_{imp}}{N_{tot}}\right)_{d} - \left(\frac{N_{imp}}{N_{tot}}\right)_{p} \,,
\end{equation}

\noindent where $N_{imp}$ is the number of particles originating from the impactor and $N_{tot}$ is the total number of particles in the post-impact planet or disk, indicated by the subscripts $p$ and $d$, respectively. A positive value of $\delta_{pd}$ therefore indicates that the protolunar disk is enriched in impactor material relative to the Earth, whereas a negative value of $\delta_{pd}$ indicates that the disk is depleted in impactor material relative to the Earth. 

The measured isotopic ratios are indistinguishable to within $5\sigma$. Such a sensitivity of $\delta_{pd}$ is difficult to achieve in SPH simulations, as $\delta_{pd}$ depends on the resolution of the simulation and the mass (i.e., number of particles) of the post-impact disk. Nevertheless, values of $\delta_{pd}$ near or equal to zero should be interpreted as favorable because they allow for a larger pre-impact compositional difference between the target and impactor and rely less on post-impact equilibriation processes.

\section{Methods} \label{sec:methods}

The giant impact simulations presented in this work are performed with the Smoothed Particle Hydrodynamics (SPH) code \texttt{Gasoline} \citep{wadsleyGasolineFlexibleParallel2004}. The version of \texttt{Gasoline} used in this work includes modifications as described in \citet{reinhardtNumericalAspectsGiant2017} and \citet{reinhardtBifurcationHistoryUranus2020} and uses a generalized equation of state (EOS) interface \citep{meierEOSlib2021, meierANEOSmaterial2021}. This version of \texttt{Gasoline} has been used extensively for giant impact simulations \citep{chauFormingMercuryGiant2018, reinhardtBifurcationHistoryUranus2020, timpeMachineLearningApplied2020, chauCouldUranusNeptune2021, meierEOSResolutionConspiracy2021, wooDidUranusRegular2022}.

\subsection{Equations of State (EOS)} \label{sec:method-eos}

The simulations in this work follow collisions between bodies with distinct compositional layers, namely iron cores and rocky mantles. We use the ANEOS (ANalytic Equation of State) EOS \citep{thompsonImprovementsCHARTRadiationhydrodynamic1974} to model the materials, specifically iron  \citep{emsenhuberSPHCalculationsMarsscale2018} for the core and dunite \citep{benzOriginMoonSingle1989} for the mantle. ANEOS is based on fitting analytic expressions of the Helmholtz free energy in different phases of the material to experimental data. It covers a wide range of densities and temperatures and faithfully models shock compression and release. This makes it a very popular choice for impact simulations.

\subsection{Pre-impact models} \label{sec:method-models}

Each collision begins with two distinct bodies, designated the \textit{target} and \textit{impactor}, whereby the target is the more massive of the two bodies. The particle representation of these bodies are created with the \texttt{BALLIC} code \citep{reinhardtNumericalAspectsGiant2017}, including improvements for multi-component models as described in \citet{chauFormingMercuryGiant2018} and \citet{reinhardtBifurcationHistoryUranus2020}. In this work, the models have Earth-like compositions, with an iron core (33\% by mass) and a rocky mantle (67\%). The thermal profiles of the models are constructed to be adiabatic, with surface temperatures set to $T_{s} = 1000~K$.

\subsection{Initial conditions} \label{sec:method-ic}

The pre-impact state of each collision is defined by a set of parameters that define the geometry of the collision and the internal compositions of the target and impactor. We have chosen to use the initial total angular momentum budget ($J_0$) and asymptotic relative velocity ($v_{\infty}$)---which in turn set the asymptotic impact parameter ($b_{\infty}$)---to define the initial geometries of the collisions. This is in contrast to many previous studies, which have chosen to parameterize their collisions by the impact parameter ($b_{imp}$) and velocity at the moment of impact ($v_{imp}$). Our choice is motivated by the fact that the target and impactor can undergo significant deformation prior to the actual impact which renders a determination of the impact parameter and velocity at impact problematic. To avoid any confusion, a detailed definition of the asymptotic parameters ($v_{\infty}$ and $b_{\infty}$) and their relation to the parameters at the moment of impact ($v_{imp}$ and $b_{imp}$), as well as to the initial positions of the colliding bodies at the start of each simulation ($v_{ini}$ and $b_{ini}$), is provided in Appendix~\ref{sec:appendix-parameters}.

The total mass of the colliding bodies in this study is always $M_{tot} = M_{targ} + M_{imp} = 1.05~M_{\oplus}$. Given the typical post-impact disk masses resulting from the simulations considered here, this is approximately the total mass required to produce a $1~M_{\oplus}$ Earth and a protolunar disk with a favorable mass. The masses of the target and impactor for given $M_{tot}$ and $\gamma$ are then,

\begin{equation}
M_{targ} = M_{tot} \left( \frac{1}{\gamma + 1} \right)\,,\label{eq:target-mass}
\end{equation}

\begin{equation}
M_{imp} = M_{tot} \left( \frac{\gamma}{\gamma + 1} \right)\,.\label{eq:impactor-mass}
\end{equation}

\noindent where $\gamma$ is the impactor-to-target mass ratio. 

The fundamental parameters that we vary between simulations are then the impactor-to-target mass ratio ($\gamma$), the initial total angular momentum budget ($J_{0}$), and the asymptotic relative velocity ($v_{\infty}$). Given the three parameters above, the asymptotic impact parameter ($b_{\infty}$) is calculated as follows:

\begin{equation}
b_{\infty} = \frac{J_{0}}{M_{tot} v_{\infty}} \frac{\left( \gamma + 1 \right)^{2}}{\gamma} \,.\label{eq:com-binf}
\end{equation}

\noindent The factor of $(\gamma + 1)^2 / \gamma$ in Equation~\ref{eq:com-binf} is required because $J_0$ is given in the center of mass frame, while $b_{\infty}$ is calculated in the target's frame of reference, and angular momentum is not conserved under such frame transitions.

Regarding the initial distance between the colliding bodies ($d_{ini}$), it is computationally prohibitive to place the target and impactor at large distances from each other. Therefore, we place the target and impactor close enough together that the pre-impact phase of the simulation is computationally tractable but far enough apart that they are not yet subject to significant mutual gravitational interactions (causing significant deformation and tidal interaction). Therefore, for all simulations in this work, $d_{ini} = 10~R_{crit}$, where $R_{crit} = R_{targ} + R_{imp}$ and $R_{targ}$ and $R_{imp}$ are the radii of the (non-rotating) target and impactor, respectively.

\begin{figure*}[ht!]
\plotone{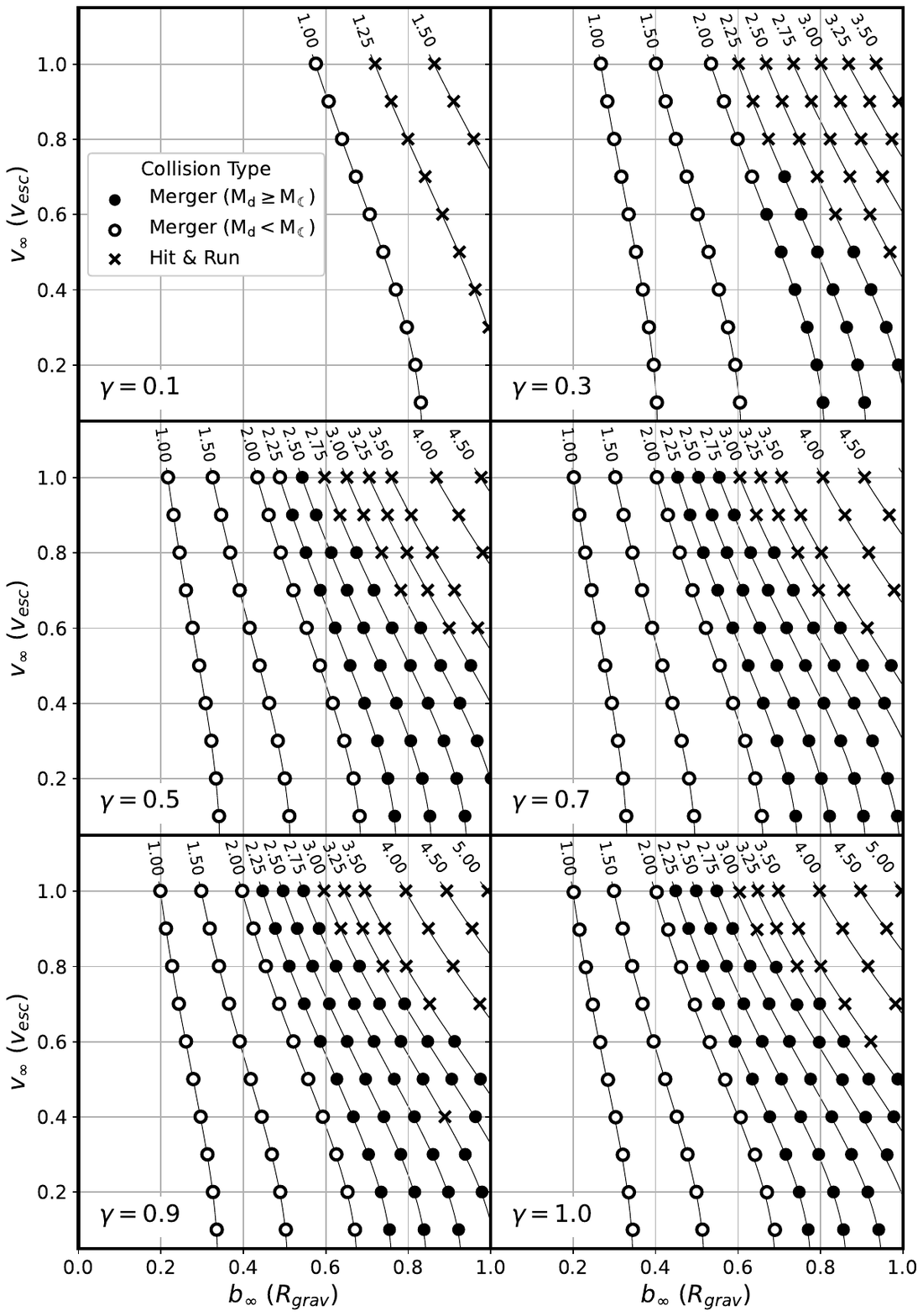}
\caption{Pre-impact initial conditions for collisions with non-rotating targets and impactors. The pre-impact trajectory is defined by specifying the initial total angular momentum budget ($J_0$) and the asymptotic relative velocity ($v_{\infty}$). The asymptotic impact parameter ($b_{\infty}$) is then computed according to Equation \ref{eq:com-binf}. Each line tracks a constant $J_0$, where the numbers at the top of each line specify the value of $J_0$. \label{fig:figure1}}
\end{figure*}

We reiterate that in Paper I we only consider impact scenarios without pre-impact rotation of the target or impactor. In Paper II, we explore pre-impact rotation of the target and impactor as variable parameters. The initial conditions of the collisions simulated in this work are shown in Figure~\ref{fig:figure1}.

\subsection{Post-impact analysis} \label{sec:analysis}

In order to classify collisions into distinct outcomes, we use the \texttt{SKID} group finder \citep{stadelCosmologicalNbodySimulations2001} to identify the number and mass of post-impact fragments. \texttt{SKID} identifies coherent, gravitationally bound clumps of material. It does this by identifying regions that are bounded by a critical surface in the density gradient and then removes the least bound particles one by one from the resulting structure until all particles are self-bound. The clumps identified by \texttt{skid} are then combined if they are co-located. 

For collisions where at least one surviving post-impact body is identified by SKID, an analysis is carried out using \texttt{pynbody} \citep{pontzenPynbodyNBodySPH2013}, a Python package for analyzing astrophysical SPH simulations. As a first step, we identify the largest remnant (LR), which corresponds to the surviving target, the second largest remnant (SLR), which generally corresponds to the surviving impactor, and the ejecta, which corresponds to particles that are gravitationally unbound from the post-impact remnant(s). Once the LR and---if it exists---SLR are identified, we classify the collisions by outcome.

\subsubsection{Collision outcomes} \label{sec:collision-outcomes}

A diverse range of outcomes are possible for pairwise collisions. However, in the range of pre-impact conditions likely to lead to the formation of the Moon, there are only three types of outcomes that are relevant:

\begin{description}

    \item[Merger] The impactor merges with the target as a result of the initial impact. Some fraction of the colliding material will be lost as ejecta, but this fraction is generally small.

    \item[Hit \& run] The impactor survives the initial impact and has enough energy to escape the gravitational pull of the target. In this work, we only analyze the post-impact state of collisions that have been classified as a merger. We therefore ignore hit \& run cases but note that, in theory, both of the post-impact remnants in such a scenario could host a circumplanetary disk.
    
    \item[Graze \& merge] The impactor survives the initial impact but does not have enough energy to escape the gravitational pull of the target. However, note that if the bound impactor's orbit takes it beyond the Hill radius of the Earth ($r_{apo} > R_{Hill}$), then we consider the collision to be a hit \& run. The surviving impactor will therefore re-impact the target at a later time. In these cases, we continue to run the simulation forward until the re-impact has occurred and the collision has resolved. Once these collisions have resolved, their outcomes are re-classified. 
    
\end{description}

The collisions in this study are classified according to one of these categories. Note that in Paper I, which explores only non-rotating collisions, no graze \& merge outcomes were found. However, graze \& merge do result in Paper II and we therefore include their definition above for completeness.

\subsubsection{Disk finder} \label{sec:disk-finder}

In order to identify the planet (i.e., Earth), protolunar disk, and ejecta following the impact, we employ a novel disk finding algorithm. This disk finding algorithm differs from previous approaches in that it determines the planet's radius ($R_{p}$) by finding the radius at which the median rotation rate of local particles deviates significantly from the expected solid-body rotation rate. In a subsequent step, the algorithm largely follows previous approaches by assigning particles exterior to $R_p$ to either the planet, disk, or ejecta based on the periapsis distance of each particle's orbit. This disk finding algorithm is described in detail in Appendix \ref{sec:appendix-disk-finder}. 

\section{Results \& Discussion} \label{sec:results}

In this paper (Paper I), we present the results of pairwise collisions between non-rotating targets and impactors. The set of collisions in this work consists of two distinct subsets: the main subset consists of 435 impacts by relatively large ($0.1 \leq \gamma \leq 1.0$), low-velocity ($v_{\infty} \leq v_{esc}$) impactors. The smaller subset consists of 62 impacts by relatively small ($0.025 \leq \gamma \leq 0.05$), high-velocity ($1.2~v_{esc} \leq v_{\infty} \leq 3~v_{esc}$) impactors. Of the 497 collisions simulated in total, the following outcomes are observed: 355 mergers and 142 hit \& runs. In the results that follow, we only consider the 355 collisions that resulted in a single large post-impact body (i.e., the mergers). The collisions in this work which have been classified as hit \& run are not relevant for lunar formation because the mass of the resulting planet is significantly lower than the mass of Earth. The hit \& run cases are therefore not considered in the results that follow.

The distributions of collision outcomes for the large subset is shown in Figures \ref{fig:figure1}. As would be expected, hit \& run collisions result from grazing impacts with relatively high velocities (the top right region in each panel). Circles indicate collisions that resulted in a merger, with filled circles representing collisions that generated a protolunar disk of at least one lunar mass ($M_d \geq M_{\leftmoon}$) and open circles representing collisions that generated either no disk or a disk with less than one lunar mass ($M_d < M_{\leftmoon}$). 

The collision outcomes for the small subset of low-mass, high-velocity impactors are not shown, as all of the collisions in this subset failed to produce sufficiently massive protolunar disks. Indeed, the most massive disk produced by these collisions is less than 1\% of the lunar mass ($M_d < 0.01~M_{\leftmoon}$). Therefore, we rule out this class of collisions and ignore the associated simulations in the results and discussion that follow.

\subsection{Protolunar disk mass} \label{subsec:results-disk-mass}

\begin{figure*}[h!]
\plotone{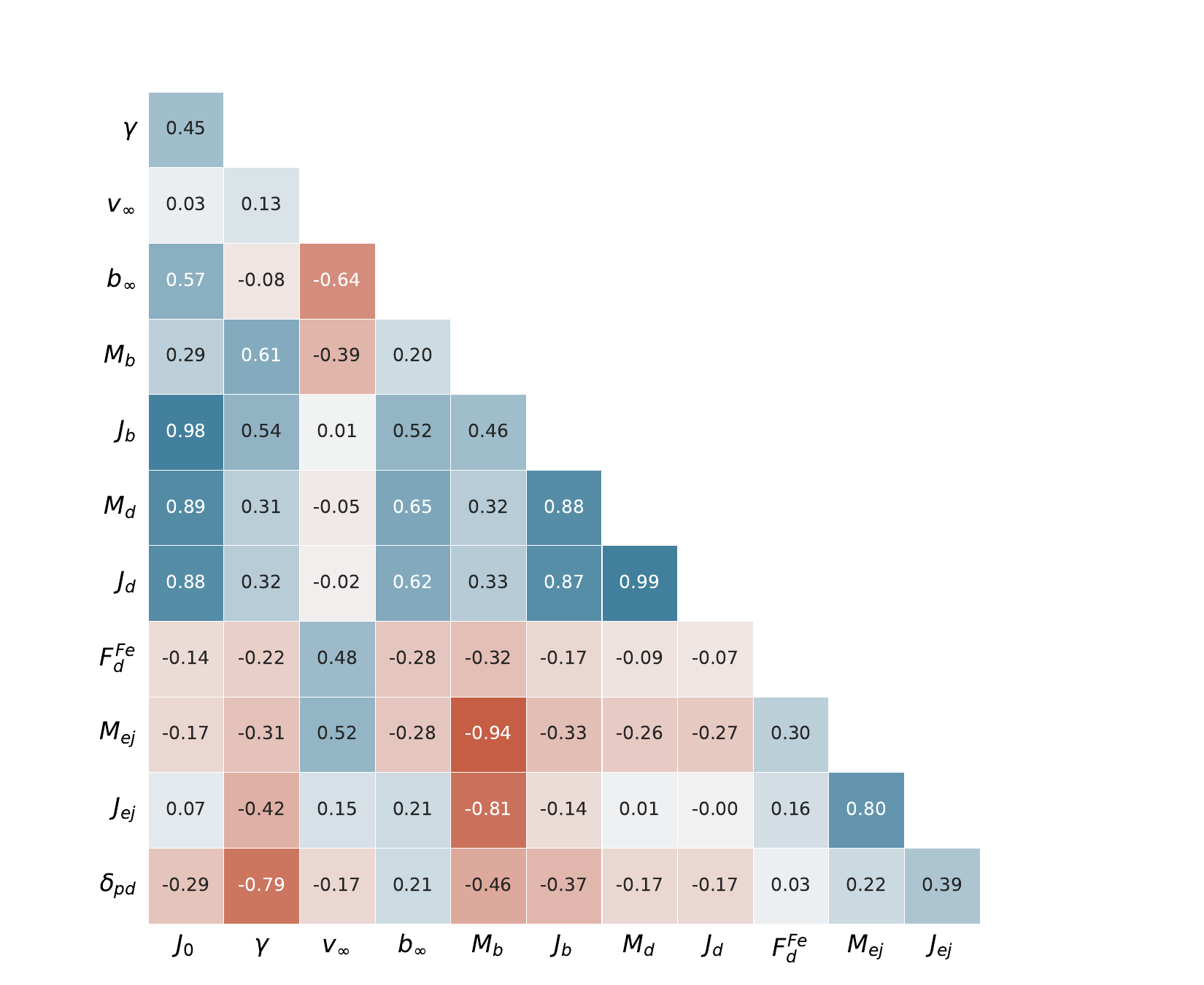}
\caption{Pearson correlation coefficients for a selection of pre-impact and post-impact properties. Blue squares indicate a positive correlation between properties, with stronger correlations marked by darker shades of blue. Red squares indicate a negative correlation, with darker shades of red indicating a stronger correlation. \label{fig:pearson_corr}}
\end{figure*}

We find 179 mergers in our dataset that produce protolunar disks of at least one lunar mass. Figure \ref{fig:figure1} illustrates that significantly more angular momentum than the current budget of the Earth-Moon system is required to generate a protolunar disk with at least one lunar mass. The demarcation of mergers with and without a sufficiently massive protolunar disk hints at a strong relationship between the pre-impact angular momentum budget ($J_0$) and post-impact protolunar disk mass ($M_d$). Indeed, the Pearson correlation coefficients ($r$) measured between the pre- and post-impact properties (see Figure \ref{fig:pearson_corr}) indicate that $J_0$ is by far the strongest determinate of $M_d$, with larger pre-impact angular momentum budgets driving more massive disks (Pearson $r = 0.88)$. The impactor-to-target mass ratio ($\gamma$) also plays a significant role in determining $M_d$, with higher mass ratios resulting in more massive disks (Pearson $r = 0.36$).

\begin{figure*}[ht!]
\plotone{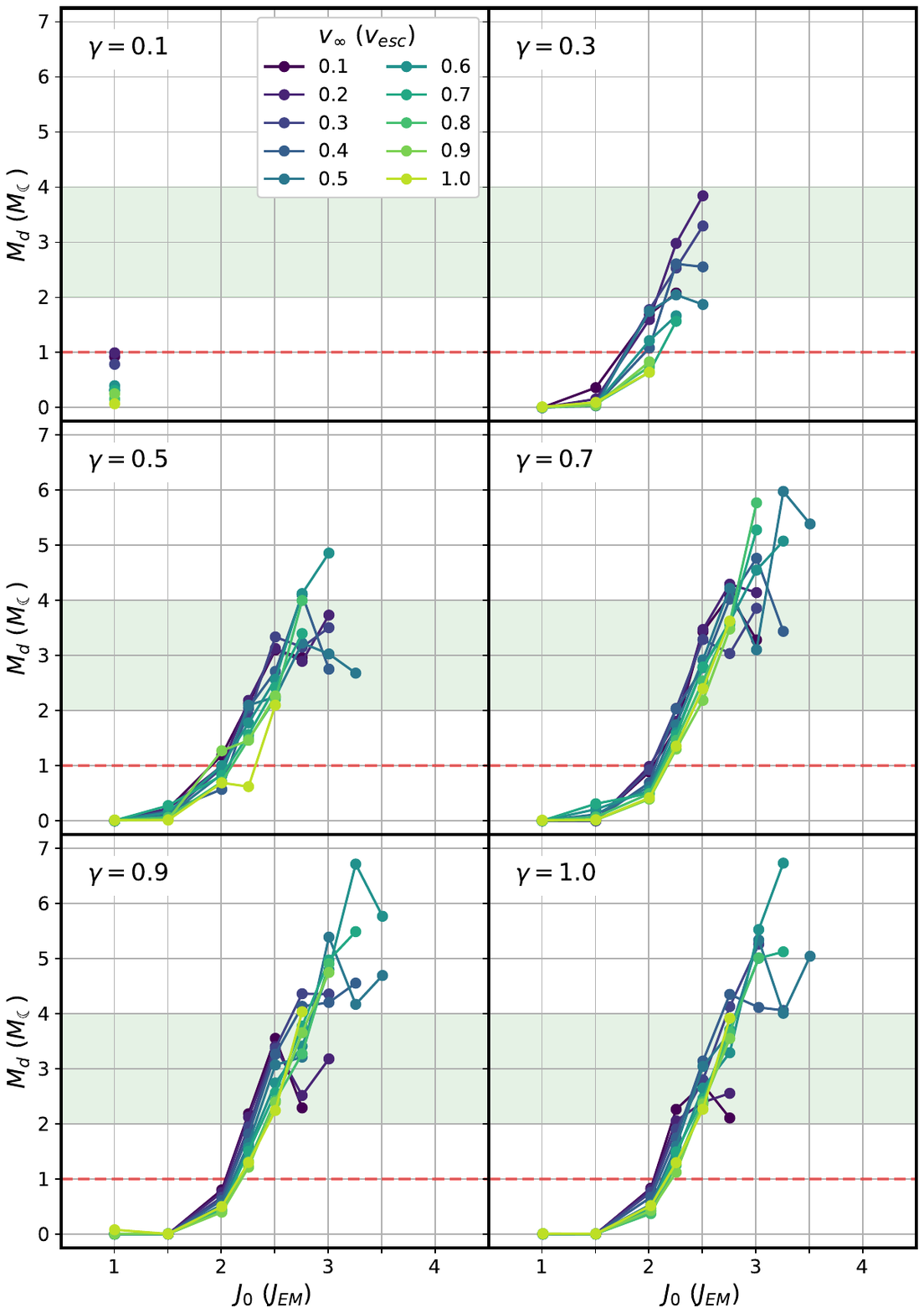}
\caption{Post-impact circumplanetary disk mass ($M_d$) for collisions between non-rotating bodies. The disk mass is shown as a function of the pre-impact angular momentum budget ($J_{0}$). Each panel presents a distinct impactor-to-target mass ratio ($\gamma$). The red dashed horizontal line is the absolute minimum disk mass ($M_d \geq M_{\leftmoon}$) needed to form the Moon assuming perfect accretion from the protolunar disk. The green-shaded region constrains the range of disk masses suggested by post-impact N-body simulations of accretion in the circumplanetary disk. \label{fig:figure2a}}
\end{figure*}

Figure \ref{fig:figure2a} most clearly illustrates the relationship between $J_0$ and $M_d$. From this relationship, it is clear that in order to generate a disk with enough mass to form the Moon ($M_d \geq M_{\leftmoon}$), a pre-impact angular momentum budget of at least $J_0 \simeq 2~J_{EM}$ is required. Note that a protolunar disk mass of $M_d = M_{\leftmoon}$ implies a 100\% accretion rate during the subsequent accretion of the Moon from the disk. Such an accretion rate is unrealistic and therefore disk masses will need to be significantly higher in order to provide enough material to form a lunar-mass object under the assumption of imperfect accretion. Indeed, N-body studies of lunar accretion from circumplanetary disks suggest that realistic accretion rates are closer to 25-50\% \citep{kokuboEvolutionCircumterrestrialDisk2000}. Under these constraints, a disk mass of $M_d \geq 2~M_{\leftmoon}$ is required, suggesting that the minimum viable pre-impact angular momentum budget is closer to $J_0 \simeq 2.25~J_{EM}$.

In the context of collisions between non-rotating bodies, this result presents significant difficulties for the giant impact hypothesis, as it implies that a post-impact process capable of removing more than $J_{EM}$ must exist. Currently, the only known process by which a significant amount of angular momentum can be removed from the Earth-Moon system following an impact is the Solar Evection Resonance (SER). However, it is still unclear how much angular momentum an SER could have removed from the Earth-Moon system under realistic conditions, with estimates varying from a few percent \citep{tianCoupledOrbitalthermalEvolution2017} to several $J_{EM}$ \citep{cukMakingMoonFastSpinning2012} depending on the underlying tidal model. This result leaves the lunar formation community with two distinct---but certainly not mutually exclusive---potential solutions for rescuing the giant impact hypothesis. One solution would be to demonstrate a sufficiently effective SER. The existence of such an SER is beyond the scope of this work, but we note that further research is needed to understand this process. Another solution may be realized by allowing for rapid pre-impact rotation of the target and impactor. We explore whether or not such pre-impact rotation can reconcile the angular momentum problem in Paper II and reserve the effectiveness SER for future work.

\subsection{Composition of the protolunar disk} \label{subsec:disk-composition}

Under the giant impact hypothesis, the Moon is assumed to have accreted from the circumplanetary disk created by the impact. Thus, the composition of the Moon is largely determined by the composition of the post-impact protolunar disk. Two compositional constraints are relevant in this respect: the iron fraction of the disk ($F^{Fe}_{d}$) and the fraction of disk material originating from the impactor body. The latter constraint is important in relation to the composition of the post-impact Earth; in order to explain the isotopic similarities between the Earth and the Moon, the fraction of impactor material in the post-impact Earth and protolunar disk should be similar.

\subsubsection{Iron content} \label{subsubsec:iron-content}

\begin{figure*}[ht!]
\plotone{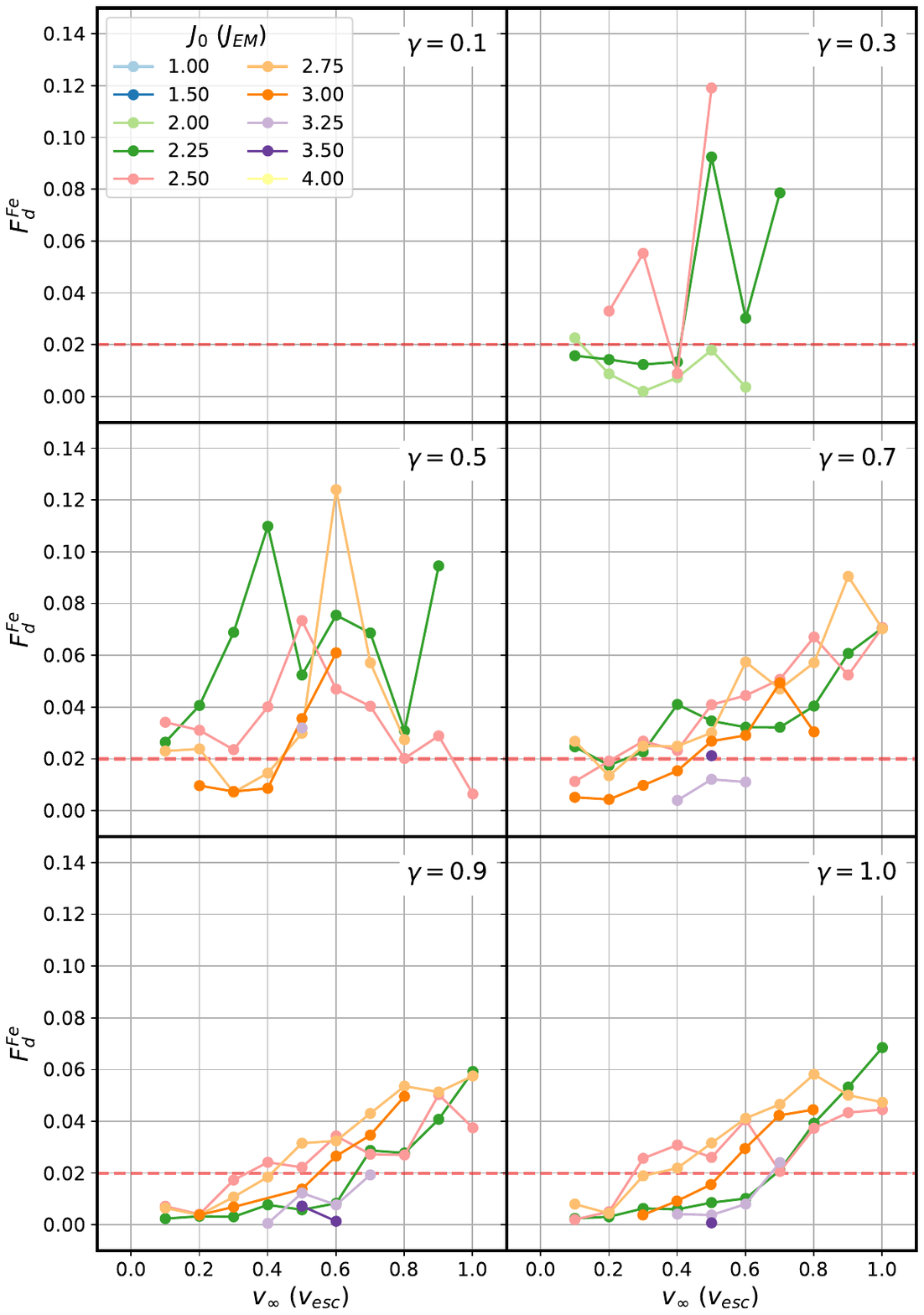}
\caption{Post-impact iron fraction of the protolunar disk ($F^{Fe}_{d}$) for disks with at least one lunar mass, shown as a function of the asymptotic relative velocity ($v_{\infty}$). The iron fraction of the Moon has been constrained at no more than 2\% by mass, indicated by the red dashed horizontal line. This line represents the maximum iron fraction if we assume that iron is accreted from the circumplanetary disk into the Moon at the same rate as silicates. This constraint can be relaxed if we assume that iron is accreted into the growing Moon less efficiently than silicates. \label{fig:figureFeDisk}}
\end{figure*}

A successful simulation should avoid injecting too much iron into the protolunar disk. While the iron fraction of the protolunar disk should preferably be less than 2\%, the allowable iron fraction can be increased if we assume that iron is accreted into the Moon less efficiently than mantle material. Figure \ref{fig:figureFeDisk} illustrates a strong relationship between the asymptotic relative velocity ($v_{\infty}$) and $F^{Fe}_{d}$ for $\gamma \gtrsim 0.5$, while for low-$\gamma$ the fraction of iron in the disk is difficult to predict. This latter uncertainty appears to be due to the tendency of low-$\gamma$ impacts to produce relatively large intact fragments. The iron fraction of these fragments and their subsequent inclusion in the protolunar does not depend predictably on the pre-impact parameters. In future work, it will be important to study the behavior of these fragments at much higher numerical resolutions. 

These results indicate that high-$\gamma$ ($\gamma \gtrsim 0.5$), low-velocity ($v_{\infty} < 0.7~v_{esc}$) impacts are favored in order to keep the iron fraction of the disk sufficiently low. It is possible that higher disk iron fractions could be successful, but this implies a spatial distribution of the iron in the disk which prevents it from being accreted at the same rate as mantle material. The long-term behavior of accretion is beyond the scope of this work and the maximum disk iron fraction should be constrained by future post-impact accretion studies. However, it is reasonable to expect that realistic disk iron fractions would not be more than a few percent.

\subsubsection{Isotopic composition} \label{subsubsec:isotopic-composition}

\begin{figure*}[ht!]
\plotone{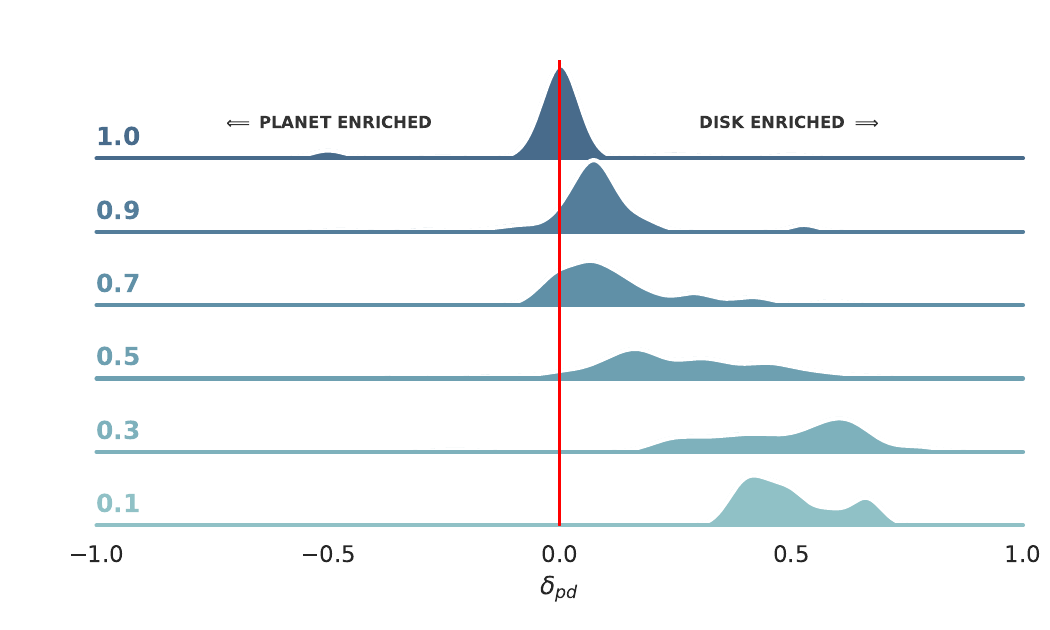}
\caption{Distribution of compositional variations between the post-impact Earth and protolunar disk, where $\delta_{pd}$ is determined according to Equation \ref{eq:delta-mix}. A value of zero indicates that the planet and disk are composed of the same fraction of impactor material. A value greater than zero indicates that the disk is enriched in impactor material relative to the planet, and vice versa. The numbers on the left are the impactor-to-target mass ratio ($\gamma$). Note that as $\gamma$ decreases, the disk tends to be composed predominantly of impactor material. \label{fig:figure_mixing}}
\end{figure*}

The compositional difference between the protolunar disk and the proto-Earth is quantified by $\delta_{pd}$, which is defined in Equation~\ref{eq:delta-mix}. The strongest determinant of $\delta_{pd}$ is the impactor-to-target mass ratio ($\gamma$), with lower values of $\gamma$ resulting in increasingly large differences between the fractions of impactor material in the Earth and protolunar disk. The Pearson correlation coefficient for $\gamma$ and $\delta_{pd}$ quantifies this inverse relationship, with Pearson $r = -0.79$. Figure \ref{fig:figure_mixing} most clearly demonstrates this relationship. Overall, it is very difficult to achieve the level of compositional similarity suggested by isotopic measurements. Only at equal or very nearly equal-mass mergers ($\gamma \rightarrow 1$) does the difference in the fraction of impactor material between the Earth and protolunar disk approach zero (see Figure \ref{fig:figure_mixing}). This result strongly favors near-equal mass mergers if there is any significant compositional difference between the pre-impact target and impactor.

\subsubsection{Post-impact mixing}

If the atmosphere of the Earth and the inner edge of the protolunar disk remain in contact following the impact, then it is possible that these reservoirs could continue to exchange material. This could have the effect of reducing the iron fraction of the disk (assuming that iron is preferentially lost to the planetary atmosphere) or equilibrating the isotopic composition of the Earth and protolunar disk. In order to equilibrate the compositions of the planet and disk post-impact, processes that rely on a link between the planet's mantle (via its post-impact atmosphere) and inner disk have been suggested. 

\begin{figure*}[ht!]
\plotone{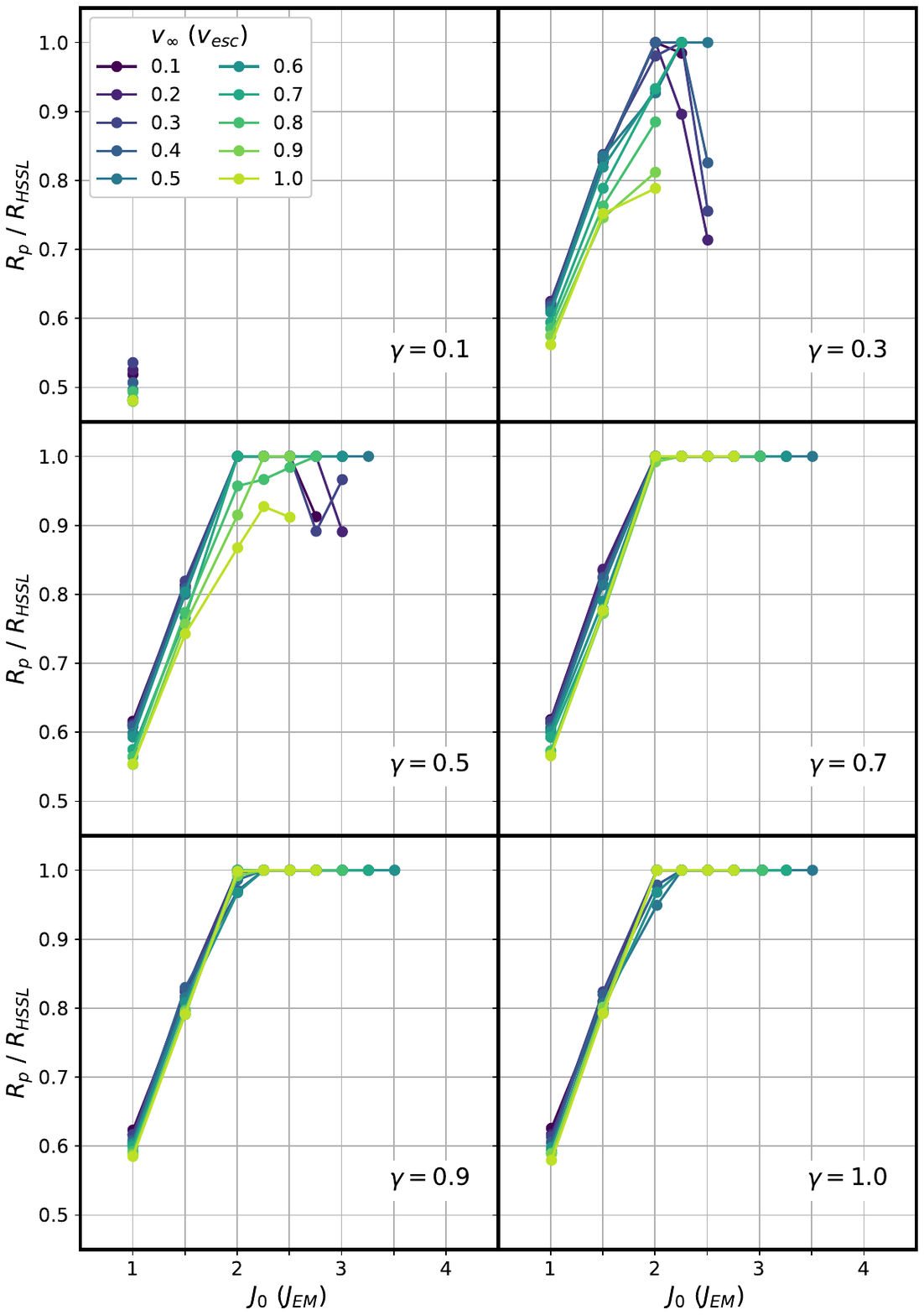}
\caption{Post-impact equatorial radius of the central body ($R_p$) relative to the Hot Spin Stability Limit (HSSL) radius ($R_{HSSL}$). A value at or near one indicates that the post-impact central body is rotating at its maximum rate and may still be actively exchanging material with the circumplanetary disk or ejecta. \label{fig:figureRhssl}}
\end{figure*}

A post-impact structure known as a \textit{synestia} is currently thought to offer such a link. However, as Figure \ref{fig:figureRhssl} shows, not all collisions will result in such a post-impact structure. Only those collisions in the hot-spin stability limit (HSSL) regime are candidates for post-impact compositional equilibration. To reach the HSSL regime, large initial angular momentum budgets are required ($J_0 \geq 2$). For lower mass ratios ($\gamma < 0.5$), too much angular momentum can prevent the post-impact system from being in the HSSL regime.

\subsection{Angular momentum budget} \label{subsec:results-angular-momentum}

The angular momentum budget of the Earth-Moon system is extremely well constrained and angular momentum is difficult to alter via post-impact processes. Therefore, a critical question for potential Moon-forming impacts is how much angular momentum remains in the bound material (i.e., the Earth and protolunar disk) following the impact. The results in this work demonstrate that very little of the pre-impact angular momentum budget ($J_0$) is lost via the impact-generated ejecta. Given that at least $J_0 \simeq 2.25 J_{EM}$ is required to generate a significantly massive protolunar disk (Figure \ref{fig:figure2a}), this implies that there must be a post-impact process capable of removing at least $\sim 1.25 J_{EM}$ if non-rotating cases are to be successful.

For collisions between non-rotating bodies, the initial total angular momentum budget ($J_{0}$) strongly determines two important post-impact quantities for lunar formation: the post-impact angular momentum budget of the bound material ($J_{b}$) and the mass of the protolunar disk ($M_{d}$). The Pearson correlation coefficients for $J_0 - J_b$ and $J_0 - M_d$ quantify these effects and are $r=0.88$ and $r=0.89$, respectively. Moreover, the protolunar disk mass almost perfectly correlated with the angular momentum budget of the protolunar disk, evincing a coefficient of $r=0.99$. For collisions that result in full or partial accretion---i.e., those that do not result in a hit \& run---almost all of the angular momentum remains with the bound material. Only at lower impactor-to-target mass ratios ($\gamma \lesssim 0.5$) is significant angular momentum carried away by the ejecta for high-velocity impacts. 

\subsection{Ejecta} \label{subsec:results-ejecta}

For high impactor-to-target mass ratios ($\gamma \gtrsim 0.5$), the mass of the ejecta never exceeds 5\% of the initial total mass ($M_{ej} < 0.05~M_{tot}$). For collisions with $\gamma > 0.5$, the ejecta mass generally decreases with an increasing angular momentum budget. However, for collisions with $\gamma < 0.5$, the trend is reversed and the ejecta mass increases with an increasing angular momentum budget.

Similarly, the fraction of angular momentum carried away with the ejecta is generally small. The exception to this is at low-$\gamma$, where higher-velocity impacts start to produce ejecta that carries away a significant fraction of the initial total angular momentum. These low-$\gamma$, high-velocity cases correspond to the cases with relatively large ejecta masses ($\sim$ 5\%).

\subsection{Promising classes of impacts} \label{subsec:results-promising-impacts}

The results presented here show that non-rotating collisions cannot generate sufficiently massive protolunar disks below $J_0 \simeq 2 J_{EM}$. If the angular momentum constraint is relaxed (i.e., by assuming there exists a post-impact process that is capable of removing large amounts of angular momentum from the system), then a handful of collisions in our dataset are capable of meeting the remaining constraints. We consider two different sets of constraints: one strict and one more permissive. 

In the permissive case, we ask which simulations produce a disk of at least one lunar mass ($M_d \geq M_{\leftmoon}$) with an iron fraction of less than 4\% ($F^{Fe}_{d} \leq 0.04$). This assumes an overall accretion efficiency of 50-100\% in the protolunar disk and an iron accretion efficiency of $<50\%$. In the strict case, we ask which simulations produced a disk within the mass range suggested by accretion studies ($2M_{\leftmoon} \leq M_d \leq 4M_{\leftmoon}$) with a disk iron fraction below 2\% ($F^{Fe}_{d} \leq 0.02$) and post-impact compositional difference between the proto-Earth and protolunar disk of less than 5\% ($\left| \delta_{pd} \right| \leq 0.05$). 

Two facts conspire to rule out low-gamma ($\gamma < 0.2$) collisions between non-rotating bodies as viable lunar formation scenarios. First, our results demonstrate that a minimum pre-impact angular momentum budget of $2~J_{EM}$ is required to produce a sufficiently massive disk. Second, for $\gamma < 0.2$, there are no valid trajectories resulting in collisions for $J_{0} > 2~J_{EM}$. This raises interesting questions for the non-rotating canonical Moon-forming impact because, given the results presented here, such an impact cannot produce a sufficiently massive protolunar disk and could therefore not have led to the formation of the Moon.

\begin{figure*}[ht!]
\plotone{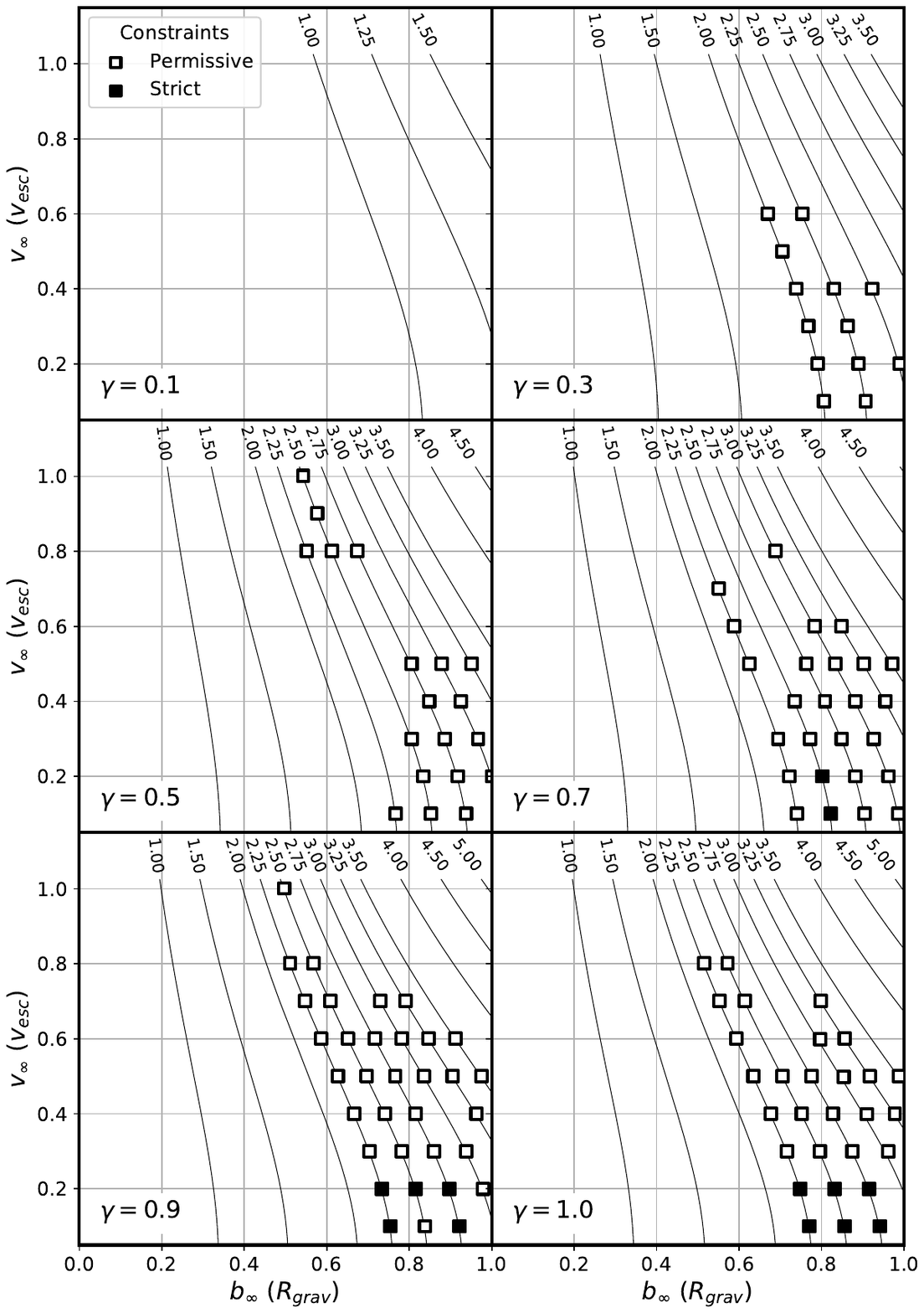}
\caption{Successful cases according to a set of permissive ($M_d \geq M_{\leftmoon}$ and $F^{Fe}_d \leq 0.04$) and strict ($2~M_{\leftmoon} \leq M_d \leq 4~M_{\leftmoon}$, $F^{Fe}_{d} \leq 0.02$, and $|\delta_{pd}| \leq 0.05$) constraints. These results show that non-rotating collisions cannot generate favorable disk properties while simultaneously reproducing the current Earth-Moon angular momentum budget. If the angular momentum constraint is relaxed, then only a handful of collisions can meet the remaining constraints, including a sufficiently massive protolunar disk mass, a disk iron fraction of less than 2\%, and a difference in post-impact composition of the Earth and protolunar disk of less than 5\%. The distribution of these cases suggests that the most promising case of impacts are low-velocity, high-angular momentum impacts between near-equal mass bodies. \label{fig:figure-best-cases}}
\end{figure*}

While none of the collisions in our dataset are able to reconcile the angular momentum constraint, some collisions are more favorable in terms of post-impact compositional constraints. Indeed, collisions between near-equal mass bodies ($\gamma \simeq 1$) produce Earths and protolunar disks with nearly indistinguishable isotopic compositions ($F^{imp}_{p} \simeq F^{imp}_d$). Of course, while this measure is a crude proxy for actual isotopic compositions, such values do indicate a much more favorable initial compositional difference that may more easily reach equilibrium via post-impact mixing processes.

Taken together, the simulations in this work suggest that the most favorable impact conditions are low-velocity ($v_{\infty} < 0.7~v_{esc}$) impacts between near-equal mass bodies ($\gamma \simeq 1$) with pre-impact angular momentum budgets of $J_0 \geq 2.25~J_{EM}$. These collisions are likely to produce sufficiently massive protolunar disks with favorable compositions and iron fractions. We note that this class of impacts (i.e., low-velocity, equal-mass mergers) most closely corresponds to the Moon-forming impacts proposed by \citet{canupFormingMoonEarthlike2012}. The results here necessitate, however, a post-impact process capable of removing at least the equivalent of the current angular momentum budget of the Earth-Moon system, which lends support to the stronger class of SER proposed by \citet{cukMakingMoonFastSpinning2012}. 

\section{Conclusions} \label{sec:conclusions}

We simulate 497 pairwise collisions between differentiated non-rotating bodies. Two distinct sets of collisions are considered: a main set of 435 collisions with large impactor-to-target mass ratios ($0.1 \leq \gamma \leq 1$) and asymptotic relative velocities equal to or below the mutual escape velocities of the colliding bodies ($v_{\infty} \leq v_{esc}$); and a smaller set of 62 collisions with small mass ratios ($0.025 \leq \gamma \leq 0.05$) and velocities above the mutual escape velocities of the colliding bodies ($1.2~v_{esc} \leq v_{\infty} \leq 3~v_{esc}$). We reiterate that the conclusions presented here assume the absence of any pre-impact rotation. The effects of such rotation is the focus of Paper II.

We find that the smaller set of low-$\gamma$, high-velocity collisions between non-rotating bodies fails to produce protolunar disks with sufficient mass to explain lunar formation. Indeed, this class of collisions is unable to generate disks with more than 1\% of a lunar mass. We therefore rule out this class of impacts as candidates for Moon-forming impacts. 

In the main set of higher-$\gamma$, lower-velocity collisions, only those collisions with pre-impact angular momentum budgets of $J_0 \geq 2~J_{EM}$ are able to produce disks with the minimum viable mass budget of one lunar mass ($M_d \geq M_{\leftmoon}$). If disk mass constraints suggested by post-impact N-body accretion studies are used ($2M_{\leftmoon} \leq M_d \leq 4M_{\leftmoon}$), then only collisions with $J_0 \geq 2.25~J_{EM}$ remain as viable candidates. In the absence of pre-impact rotation, this result implies that in order to reproduce the observed angular momentum budget of the Earth-Moon system, post-impact processes that are capable of removing at least $1-1.25~J_{EM}$ must exist, which supports a strong SER mechanism.

Favorable protolunar disk compositions are only achieved by low-velocity impacts between near-equal mass bodies. Indeed, in order to avoid injecting too much iron into the protolunar disk, low-velocity ($v_{\infty} < 0.7~v_{esc}$) impacts are favored, while only near-equal mass collisions ($\gamma \to 1$) are able to produce an Earth and protolunar disk with similar isotopic compositions ($\delta_{pd} \to 0$). Differences in post-impact isotopic compositions quickly increase as $\gamma$ decreases.

Taken together, these results cast doubt on the canonical Moon-forming impact and suggest that low-velocity, high-angular momentum impacts between near-equal mass bodies \citep[e.g.,][]{canupFormingMoonEarthlike2012} are more favorable candidates for Moon-forming impacts. This class of impacts requires a process capable of removing large amounts of angular momentum from the post-impact system.

The main results of our systematic survey of potential Moon-forming impacts between non-rotating bodies are summarized as follows:

\begin{itemize}

    \item The canonical Moon-forming impact cannot generate a sufficiently massive protolunar disk to explain the Moon.

    \item For all collisions, the protolunar disk mass is strongly dependent on the initial angular momentum budget. In order to generate a protolunar disk mass of at least one lunar mass, a pre-impact angular momentum budget of at least $2~J_{EM}$ is required. This implies that a post-impact process capable of removing at least $J_{EM}$ must exist.

    \item For $\gamma \gtrsim 0.5$, the iron fraction of the protolunar disk is strongly dependent on the impact velocity ($v_{\infty}$). Therefore, low-velocity ($v_{\infty} < 0.7~v_{esc}$) grazing impacts are favored to avoid injecting too much iron into the protolunar disk.

    \item Only near-equal mass collisions ($\gamma \sim 1$) are able to produce an Earth and protolunar disk with similar compositions ($\left| \delta_{pd} \right| \sim 0$) regardless of the initial compositions of the target and impactor.
    
    \item Taken together, the results of our survey favor low-velocity, near-equal mass collisions to explain the origin of the Moon.
 
\end{itemize}

In Paper II, we systematically study whether or not pre-impact rotation can lower the amount of angular momentum required to generate sufficiently massive disks while simultaneously reproducing the other observational constraints.

\acknowledgments
We thank the anonymous reviewer for valuable suggestions and comments that helped to substantially improve the paper. This work has been carried out within the framework of the National Centre of Competence in Research PlanetS supported by the Swiss National Science Foundation under grants 51NF40\_182901 and 51NF40\_205606. The authors acknowledge the financial support of the SNSF. We acknowledge access to Piz Daint and Eiger@Alps at the Swiss National Supercomputing Centre, Switzerland under the University of Zurich's share with the project ID UZH4.


\vspace{5mm}
\facilities{Swiss National Supercomputing Centre (Piz Daint, Eiger@Alps)}


\software{Gasoline \citep{wadsleyGasolineFlexibleParallel2004,reinhardtNumericalAspectsGiant2017},
          ballic \citep{reinhardtNumericalAspectsGiant2017},
          eoslib \citep{meierEOSlib2021,meierANEOSmaterial2021},
          skid\footnote{\url{https://github.com/N-BodyShop/skid}},
          numpy \citep{harrisArrayProgrammingNumPy2020},
          scipy \citep{virtanenSciPyFundamentalAlgorithms2020},
          matplotlib \citep{hunterMatplotlib2DGraphics2007},
          pynbody \citep{pontzenPynbodyNBodySPH2013},
          GNU parallel \citep{tangeGNUParallelCommandline2011}
          }



\appendix

\section{Asymptotic Parameters} \label{sec:appendix-parameters}

In contrast to most previous studies of giant impacts, we define the initial conditions of our simulations by specifying the ``asymptotic relative velocity'' ($v_{\infty}$) and the initial total angular momentum budget ($J_0$), which in turn fix the ``asymptotic impact parameter'' ($b_{\infty}$). Asymptotic refers to an ``infinite'' initial separation which, in practice, means a distance whereby mutual gravitational interactions between the target and impactor have not yet had a chance to significantly modify the pre-impact trajectory of the bodies, nor their internal structure or rotation rates through tidal interactions. In the simulations presented in this work, an initial separation of $10~R_{crit}$ is sufficient for this purpose.

Given that many previous studies on pairwise collisions have defined the initial conditions of their simulations using the relative velocity and impact parameter at the moment of impact ($v_{imp}$ and $b_{imp}$, respectively), we provide analytic prescriptions for relating the values at impact to the asymptotic values used in this work. Note, however, that these analytic relations assume no tidal interactions between the target and impactor prior to the moment of impact (i.e., the bodies maintain perfectly spherical shapes). In SPH simulations, as in reality, this is not a valid assumption and the colliding bodies can undergo significant deformation prior to impact. Thus, we urge the reader to interpret the results of these relations with caution.

Given an asymptotic relative velocity $v_{\infty}$, we can calculate the eventual velocity at impact $v_{imp}$ (under the assumption that the target and impactor are not subject to deformation and therefore no orbital energy is lost due to tidal interactions) as follows,

\begin{equation}
v_{imp} = \left(v_{\infty}^2 + \frac{2GM_{targ}}{R_{crit}} \right)^{0.5}\,, \label{appendix_eq:vinf2vimp}
\end{equation}

\noindent where $G$ is the gravitational constant and $R_{crit} = R_{targ} + R_{imp}$ is the sum of the \emph{non-rotating} equatorial radii of the target and impactor. The use of the non-rotating equatorial radii is purely a matter of convention, but we note that it greatly simplifies the problem once arbitrary orientations of rotating bodies is involved. The associated impact parameter at the moment of impact (again assuming no deformation and no tidal effects) is then,

\begin{equation}
    b_{imp} = b_{\infty} \frac{v_{\infty}}{v_{imp}} \,, \label{eq:binf2bimp}
\end{equation}

\noindent where $v_{imp}$ is calculated as in Equation \ref{appendix_eq:vinf2vimp}.

Note that when setting up a collision, the asymptotic values are converted to the associated parameters ($v_{ini}$ and $b_{ini}$) at the distances specified by the initial separation parameter $d_{ini}$. In order to convert to these values, we follow the same approach as above and calculate $v_{ini}$ as follows:

\begin{equation}
v_{ini} = \left(v_{\infty}^2 + \frac{2GM_{targ}}{d_{ini}} \right)^{0.5}\,, \label{appendix_eq:vinf2vini}
\end{equation}

\noindent where $d_{ini}$ is set to $10 ~R_{crit}$ in this work and the other parameters are the same as in Equation \ref{appendix_eq:vinf2vimp}. Similarly, the impact parameter at the start of the simulation ($b_{ini}$) is calculated as follows:

\begin{equation}
    b_{ini} = b_{\infty} \frac{v_{\infty}}{v_{ini}} \,,\label{appendix_eq:binf2bini}
\end{equation}

\noindent where $v_{ini}$ is calculated as in Equation \ref{appendix_eq:vinf2vini}.

\section{Disk Finder} \label{sec:appendix-disk-finder}

In order to assess post-impact properties following an impact, we require an algorithm to first distinguish between the post-impact planet, circumplanetary disk, and ejecta. To this end, we have developed a novel disk-finding algorithm, which we describe in detail here. On a high level, our disk-finding algorithm begins by calculating the solid-body rotation rate for the planet using the densest particles. Using a sliding window, it then moves radially outward until it reaches the radius at which the median angular velocity of particles deviates significantly from this solid-body rotation rate. All particles within this radius are assigned to the planet. Similar to other disk-finding algorithms, it then calculates the orbits of the particles outside this radius to determine which particles will fall back onto the planet (and are therefore assigned to the planet) and which of the other particles belong to the disk and which particles to the ejecta. In detail, the following steps are performed by the disk finder:

\begin{enumerate}

    \item \textbf{Center and align snapshot} The simulation output (hereafter referred to as the ``snapshot'') is centered on the planet (whereby the planet is identified by assuming some minimum density cutoff) and the $\hat{z}$-axis of the analysis frame of reference is aligned with the global angular momentum vector $\hat{J}$ of the particles.

    \item \textbf{Radially bin particles} The particles in the snapshot are binned according to their radius. The bin range is defined by $R_{min}$ and $R_{max}$, where $R_{min} = 0.1~R_{\oplus}$ in this work. The purpose of $R_{min}$ is to exclude the noisy particles near the center of the planet, while $R_{max}$ is arbitrary so long as it is sufficiently large to capture any reasonable radius (e.g., $R_{max} = 5~R_{\oplus}$. By excluding particles well beyond the expected radius, the computational performance of the disk finder is greatly improved. The number of bins within this range depends on the resolution of the simulation (i.e., the number of particles in the snapshot $N_p$). In this work, $N_{bins} = \text{int}(10^{-2}~N_{p})$. 

    \item \textbf{Determine planet radius} Starting at the innermost bin, the disk finder steps outward along the bins. At each bin, the following steps are carried out:
    
    \begin{enumerate}

        \item A sliding window with length $\ell_{win}$ is defined that extends from $w_{min} = r_{bin} - \ell_{win}$ to $w_{max} = r_{bin}$, where the $r_{bin}$ is the midpoint of the current bin. In this work, $\ell_{win} = 0.15~R_{\oplus}$. 
    
        \item The median rotation rate of the particles within the current window ($\omega_{win}$) is computed. This is the expected ``solid-body'' rotation rate for the current bin.

        \item The median rotation rate of the particles within the current bin ($\omega_{bin}$) is computed.

        \item The fractional difference between $\omega_{win}$ and $\omega_{bin}$ is computed,

        \begin{equation}
            \Delta = \frac{\omega_{bin} - \omega_{win}}{\omega_{win}}\,.
        \end{equation}

        \item If the fractional difference is greater than a predefined threshold ($\Delta > \Delta_{crit}$), then the disk finder returns the midpoint of the current bin as the planet's radius ($R_p = r_{bin}$). In this work, the threshold is defined to be $\Delta_{crit} = 0.05$.
        
    \end{enumerate}
    
    If at any point three bins in a row are found to be empty of particles, then the disk finder returns the midpoint of the first empty bin as the planet's radius ($R_p$).

    \item \textbf{Kepler intercept} In some simulations, the post-impact Earth is still rotating at or beyond its rotational stability limit (i.e., $R_p \geq R_{HSSL}$). In these cases, the disk finder may overestimate the radius due to the large amounts of noise near the planet-disk transition. An additional step is therefore carried out to determine if the radius estimated by the disk finder exceeds the stability limit. The stability limit is approximated by identifying the radius at which the median transverse velocity in the sliding window ($v_{t,win}$) intercepts the Keplerian velocity ($v_{t,kep}$). If the radius estimated by the disk finder is larger than the radius at which the intercept occurs, the planet's radius is set to the radius of the intercept.

    \item \textbf{Differentiate disk and ejecta} Using the positions and velocities of the particles outside $R_{p}$, calculate the orbits of each particle. Particles with $e \ge 1$ are unbound and are assigned to the ejecta. For particles with $e < 1$, calculate their distance at periapsis $r_{peri}$. Those with $r_{peri} \le R_{p}$ are assigned to the planet. Those with $r_{peri} > R_{p}$ are assigned to the disk.
    
\end{enumerate}

\begin{figure*}[ht!]
\plotone{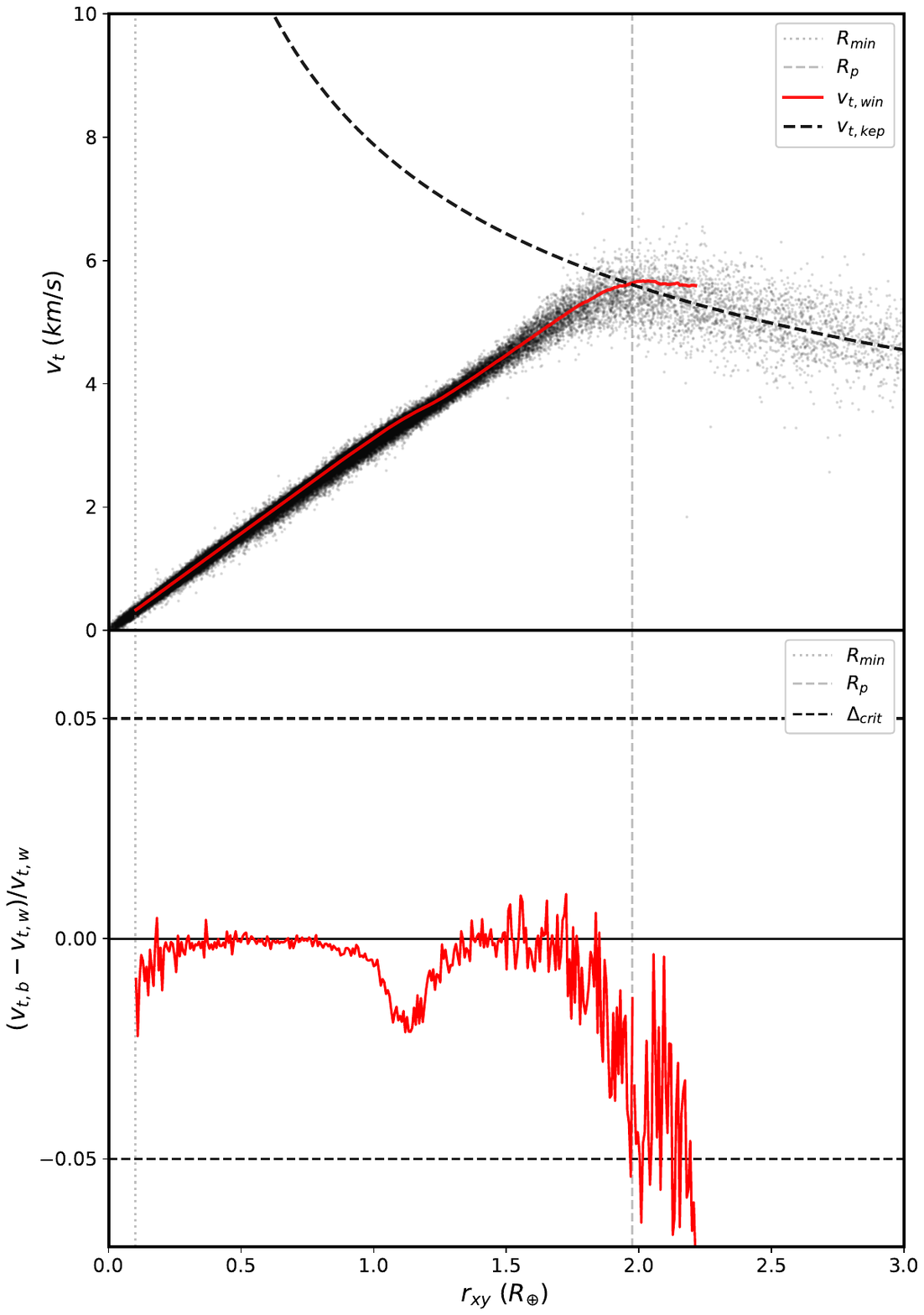}
\caption{The disk finding algorithm differentiates the post-impact planet (i.e., the proto-Earth) from the circumplanetary disk and ejecta. It identifies the radius of the planet by finding the radius at which the median rotation rate of local particles deviates significantly from the expected rotation rate (roughly corresponding to the solid-body rotation rate).\label{fig:disk-finder}}
\end{figure*}

In addition to distinguishing the post-impact structures, the disk finding algorithm can also determine the proximity of the post-impact system to the Hot Spin Stability Limit (HSSL; \citet{lockStructureTerrestrialBodies2017}). This is a useful feature of our disk-finding algorithm because it allows us to identify post-impact states where compositional mixing between the Earth's mantle and protolunar rocks can occur.

% if your bibliography is in bibtex format, use those commands:
\bibliographystyle{aasjournal} % Style BST file
\bibliography{main}

\end{document}